\documentclass[]{raa}            % referee version: for submission
\usepackage{graphicx,times}
\usepackage{natbib}

\begin{document}

   \title{Resonant cyclotron scattering
 in pulsar magnetospheres and its application to isolated neutron stars
% $^*$
%\footnotetext{\small $*$ Supported by the National Natural Science
%Foundation of China.}
}

 \volnopage{ {\bf 2009} Vol.\ {\bf 9} No. {\bf XX}, 000--000}
   \setcounter{page}{1}

   \author{Hao Tong
      \inst{1}
   \and Ren Xin Xu
      \inst{2}
   \and Qiu He Peng
      \inst{1}
   \and Li Ming Song
      \inst{3}
   }

   \institute{Department of Astronomy, Nanjing University,
               Nanjing 210093, China; {\it haotong@nju.edu.cn}\\
        \and
             Department of Astronomy, Peking University, Beijing 100871, China
        \and
             Institute of High Energy Physics, Chinese Academy of Sciences,
                Beijing 100049, China \\\vs \no
   {\small Received [year] [month] [day]; accepted [year] [month] [day] }
}

\abstract{Resonant cyclotron scattering (RCS) in pulsar
magnetospheres is considered. The photon diffusion equation
(Kompaneets equation) for RCS is derived. The photon system is
modeled three dimensionally. Numerical calculations show that there
exist not only up scattering but also down scattering of RCS,
depending on the parameter space. RCS's possible applications to the
spectra energy distributions of magnetar candidates and radio quiet
isolated neutron stars (INSs) are point out. The optical/UV excess
of INSs may caused by the down scattering of RCS. The calculations
for RX J1856.5-3754 and RX J0720.4-3125 are presented and compared
with their observational data. In our model, the INSs are proposed
to be normal neutron stars, although the quark star hypothesis is
still possible. The low pulsation amplitude of INSs is a natural
consequence in the RCS model.
  }

   \authorrunning{Tong et al. }            %author_head in even pages
   \titlerunning{RCS and its application to INSs }  % title_head in odd pages
   \maketitle

%% The author head (on even pages) and the title head (on odd pages) will be
%% automatically extracted from \author{} and \title{}. Whenever the title is too long,
%% you will be asked to supply a shorter one by inserting either \authorrunning{} or
%% \titlerunning{} before \maketitle. Anyway, you can specify your own heads.
%%
%%
%% Note: In the following text body of your manuscript, please note several differences from
%%       other major journals:
%% (1) \subsection{Please Capitalize the First Letter of Each Notional Word in Subsection Title}
%% (2) Please Capitalize the First Letter of Each Notional Word in all tables' captions

%
%________________________________________________ sections below
%

\section{Introduction}

\begin{figure}
\centering
  \includegraphics[width=0.6\textwidth]{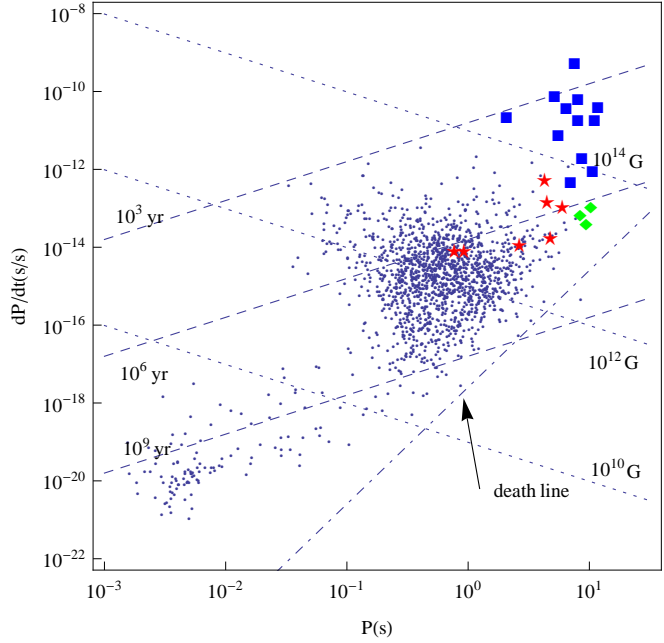}
  \caption{$P-\dot{P}$ diagram of pulsars.
  Diamonds are for INSs (Haberl 2007; Kaplan \& van Kerkwijk 2009).
  Stars are for RRATs (McLaughlin et al. 2006; McLauchlin et al. 2009).
  Magnetar (squares) and radio pulsar (dots) data are from ATNF (http://www.atnf.csiro.au/research/pulsar/psrcat/).
  The dotdashed line is the constant potential line $V=1/3\times
  10^{12}\, \mathrm{V}$.}
  \label{PPdot}
\end{figure}

Three kinds of pulsar-like objects have additionally and greatly
boosted up our konwledge about pulsar magnetospheres. They are
anomalous X-ray pulsars and soft gamma-ray repeaters (magnetars
candidates), radio quiet isolated neutron stars (INSs) (the
magnificent seven) , and rotating radio transients (RRATs). Figure
\ref{PPdot} shows their positions on the $P-\dot{P}$ diagram. Our
conventional picture of pulsar magnetospheres is provided by e.g.
Goldreich \& Julian (1969), Ruderman \& Sutherland (1975), and Cheng
et al. (1986) (for a recent review, see Kaspi et al. 2006), which is
mainly about the open field line regions (OFLRs). Few people begin
to realize that there could be interesting physics in the closed
field line regions (CFLRs) of pulsar magnetospheres. For magnetars,
it is proposed that there is strong and twisted magnetic field
around the central star (Thompson et al. 2002; Lyutikov \& Gavriil
2006). INSs are thought to be dead neutron stars, which provide a
clear specimen for magnetospheric and cooling studies (Kaspi et al.
2006; Tong \& Peng 2007; Tong et al. 2008). For RRATs, recent
modeling also indicates interesting physics in CFLRs (Luo \& Melrose
2007). The most direct evidence comes from the observations of the
double pulsar system PSR J0737-3039A/B, and there could be also
signatures of interesting physics in CFLRs of normal pulsars
(Lyutikov 2008).

The interesting physics in pulsar CFLRs are mainly related to the
plasmas there. Roughly speaking, the electron density in magnetar
CFLRs is about 4-5 orders higher than the Goldreich-Julian density
(Rea et al. 2008). In the case of RRATs, Luo \& Melrose (2007) have
proposed an idea of ``pulsar radiation belt'', like the radiation
belt of the earth. Noting the similarities between INSs and
magnetars/RRATs, we suggest that there could also be plasmas in INS
CFLRs with number density greatly higher than the Goldreich-Julian
density (i.e. the electron blanket, see Wang et al. 1998; Ruderman
2003). We observed three similarities between INSs and
magnetars/RRATs:
\begin{enumerate}
    \item Most of them are long period pulsars with spin periods about 10 seconds;
    \item They all show a non-atomic Planckian spectrum;
    \item They all have large or relatively large spin-down
    rates (indicating possible higher fields).
\end{enumerate}
With these similarities, we suggest that the physics of these three
kinds of objects should be similar, and that the very different
observational manifestations could be resulted from a different
parameter space and an evolution history.

In this paper we consider resonant cyclotron scattering (RCS)
process in pulsar magnetospheres. In section two, a brief
description of the RCS process and some basic formulas are given.
Kompaneets equation for the RCS process is deducted in  section
three. Numerical calculations are given in section four. The
application to INSs is the topic of section five. In the last two
sections, discussions and conclusions are summarized, respectively.

\section{Resonant cyclotron scattering}

Near the neutron star surface, RCS of photons is more important than
Compton scattering (Ruderman 2003). It has the following three
points.
\begin{enumerate}
    \item Given the magnetic field, the scattering
    occurs only at a specific frequency and vice versa.
    Given the frequency, the scattering occurs only at a specific
    location in the pulsar magnetosphere.
    At the resonant frequency, the cross section is about 8 orders larger
    than the Thomson cross section for typical magnetic field $10^{12}\, \mathrm{G}$.
    \item The momentum is conserved only along the z-direction (direction
    of the magnetic field).
    The field may absorb or contribute perpendicular momentum.
    \item  The particle distribution is strongly affected by the field.
    The perpendicular motion is suppressed while particles can move freely along field
    lines. Therefore the electron distribution is 1D.
\end{enumerate}
The soft X-ray spectrum of magnetars my be the result of RCS of
surface thermal emission (Rea et al. 2008). One has three ways
dealing with scattering problems (including resonant scattering).
One is solving the radiation transfer equation directly (e.g.
Lyutikov \& Gavriil 2006), the second is doing Monte Carlo
simulations (Fernandez \& Thompson 2007; Nobili et al. 2008), and
the third is introducing photon diffusion equation (Kompaneets
equation) as in the Compton scattering case (e.g. Rybicki \&
Lightman 1979). However, the Kompaneets equation for RCS has not
been developed yet. Considering it's importance in magnetar soft
X-ray emission, we present a Kompaneets equation method for the RCS
process in this paper. Also improved approximations are employed. We
find that it may account for the optical/UV excess of INSs.

Previous solution provided by Lyutikov \& Gavriil (2006) has three
problems which should be improved. The three problems are:
\begin{enumerate}
    \item It's a one dimensional treatment. Photons can only
    propagate forward or backward. All the calculations and arguments there
    are valid only in the 1D case. This will cause two
    additional problems.
    \item The angular dependence of the RCS cross section is smeared
    out. The rigorous expression is eq. (\ref{dsigma}) which we will
    discuss in the following.
    \item The down scattering of photons is dropped.
    In the 1D case, the phase space volume is proportional to $\propto
    p$. While in the 3D case it is proportional to $\propto p^3$. Noting
    that photons are bosons, the key difference between the 3D and 1D
    case is that there is no Bose-Einstein condensation in the
    later case (Pathria 2003). While the down scattering is
    Bose-Einstein condensation of photons in the low energy state (Liu et al.
    2004; Sunyaev \& Titarchuk 1980), therefore it
    can not be handled in the 1D case. Then it is not surprising
    that the authors found a net up scattering of transmitted flux.
    It is the approximations they employed that matters.
    We try to provide a 3D treatment of the photon system in this paper.
\end{enumerate}

Before going to deduction details, some basic formulas should be
given at first (You et al. 1997). Cyclotron frequency of electrons
in a given magnetic field is
\begin{eqnarray}
\nu_B &=& \frac{1}{2\pi} \frac{e B(r)}{m_e c},\label{nuB}\\
\omega_B &=& \frac{e B(r)}{m_e c},
\end{eqnarray}
where $\nu_B$ is the local cyclotron frequency, $\omega_B$ is the
angular frequency $\omega_B = 2\pi \nu_B$, $e$ is the electron
charge (absolute value), $B(r)$ is the local magnetic field, $r$ is
the distance from the point to the center of the star, $m_e$ is the
electron rest mass, and $c$ is the speed of light. When $\nu =
\nu_B$, where $\nu$ is the photon frequency, RCS occurs. The
differential cross section is
\begin{equation}\label{dsigma}
d\sigma_{\mathrm{RCS}}= \frac{3r_e c}{32} (1+\cos^2\theta)
(1+\cos^2\theta^{\prime}) \phi(\nu-\nu_B) d\Omega^{\prime},
\end{equation}
where $r_e$ is the classical electron radius, $\theta$ is the angle
between the incoming photon and the local magnetic field,
$\theta^{\prime}$ denotes the angle of the outgoing photon,
$d\Omega^{\prime}$ is the solid angle of the outgoing photon, and
$\phi(\nu-\nu_B)$ is the Lorentz line profile function, which acts
like a Dirac delta function
\begin{equation}
\phi(\nu-\nu_B) = \frac{\Gamma/4\pi^2}{(\nu-\nu_B)^2 +
(\Gamma/4\pi)^2}.
\end{equation}
Note that $\Gamma$ is the natural width
\begin{equation}
\Gamma = \frac{4e^2 \omega_B^2}{3m_e c^3}.
\end{equation}
The Lorentz line profile function has the normalization condition
\begin{equation}
\int_{-\infty}^{+\infty} \phi(\nu-\nu_B) d\nu = 1.
\end{equation}
Performing the angular integral gives the total cross section
\begin{equation}\label{RCS cross section}
\sigma_{\mathrm{RCS}} = \frac12 \pi r_e c (1+\cos^2\theta)
\phi(\nu-\nu_B),
\end{equation}
which depends on frequency.

In the case of pulsars, dipole magnetic field is always a good
approximation. The magnetic field at radius $r$ is
\begin{equation}\label{B}
B(r) = B_p \left( \frac{R}{r}\right)^3,
\end{equation}
where $B_p$ is the magnetic field at the surface of the neutron
star, and $R$ is the neutron star radius. Given the photon frequency
$\nu$, the radius at which RCS occurs is
\begin{equation}\label{rrcs}
r_{\mathrm{RCS}} = \left( \frac{\nu_{B_p}}{\nu} \right)^{1/3} R,
\end{equation}
where $\nu_{B_p}$ is the cyclotron frequency at the star surface
$\nu_{B_p} = \frac{1}{2\pi} \frac{e B_p}{m_e c}$ (only photons with
frequency smaller than $\nu_{B_p}$ will encounter RCS). For photons
in the soft X-ray band $1\, \mathrm{keV} < h\nu < 10\, \mathrm{keV}$
(with $h$ is Planck's constant), we are only considering a specific
frequency range $\nu_1 < \nu < \nu_2$. The scattering occurs in a
finite space range $r_2 < r < r_1$, where $r_2$ is the scattering
radius corresponding to frequency $\nu_2$, and $r_1$ corresponding
to frequency $\nu_1$. We assume that there are bulk of electrons
filling the space between $r_2$ and $r_1$. Beyond $r_1$ there may
also be bulk of electrons, but is is less related to the
observations in the frequency range $\nu_1$ to $\nu_2$. Finally we
introduce the optical depth of RCS
\begin{equation}\label{RCS optical depth}
\tau_{\mathrm{RCS}} = \int N_e \sigma_{\mathrm{RCS}} dr = \tau_0 (1+
\cos^2\theta),
\end{equation}
where $N_e$ is the electron number density (assuming homogenous),
\begin{equation}
\tau_0 = \frac{\pi e^2 N_e r_{\mathrm{RCS}}}{6 m_e c\,\nu}.
\end{equation}
The optical depth also depends on frequency $\propto 1/\nu^{4/3}$.
In the following sections, all optical depth are referred to their
value at the lower frequency boundary, i.e. optical depth at
$\nu_1$. During the integration of eq. (\ref{RCS optical depth}),
the spatial dependence of magnetic field (eq. (\ref{B})) is taken
into consideration. This will be used in the numerical calculation
section.

\section{Kompaneets equation for resonant cyclotron scattering}
We formulate our deduction analogous to that of Kompaneets equation
for Compton scattering (e.g. Rybicki \& Lightman 1979; You 1998;
Padmanabhan 2000).

Denote the initial and final state of the scattering as $(p_z,\
\nu,\ \vec{n})$ and $(p_z^{\prime},\ \nu^{\prime},\
\vec{n}^{\prime})$, respectively, with $p_z$ is the initial electron
momentum in the z-direction, $\nu$ the initial photon frequency,
$\vec{n}$ the propagation direction of the incoming photon, and a
prime donotes the corresponding quantity of the outgoing particles.
In strong magnetic field, electron motions perpendicular to the
magnetic field are celled into Landau energy levels. Almost all the
electrons are in the ground state (You et al. 1997). And we only
consider photons in the ground state before and after the scattering
(Herold 1979). According to energy-momentum conservation in the
non-relativistic case, we have
\begin{eqnarray}
h \nu + \frac{p_z^2}{2m_e} &=& h \nu^{\prime} + \frac{p_z^{\prime 2}}{2m_e},\\
 \left ( \frac{h\nu}{c}\cos\theta \right)  + p_z &=& \left (
\frac{h\nu^{\prime}}{c}\cos\theta^{\prime} \right) + p_z^{\prime}.
\label{conservation of energy and momentum}
\end{eqnarray}
It seems that we are dealing with a 1D distribution of electrons.

From the conservation of energy and momentum, we can calculate the
frequency change after and before the scattering $\Delta
=h(\nu^{\prime}-\nu)/k T_e$, with $k$ is Boltzmann's constant, and
$T_e$ the effect temperature of the electron system. Since we are
dealing with non-relativistic electrons $k\,T_e \ll m_e c^2$, and
consider typical photons in the X-ray band $h\nu \sim
1\,\mathrm{keV} \ll m_e c^2$, the frequency change is very small
$\Delta \ll 1$. Therefore only considering first order terms of
$\Delta$, we have
\begin{equation}\label{eq. of Delta (1st term only)}
\Delta = -\frac{x p_z}{m_e c} (\cos\theta-\cos\theta^{\prime}),
\end{equation}
where $x$ is the dimensionless frequency $x = h\nu/k T_e$. The above
expression is accurate to an order of $O(\frac{h\nu}{m_e
c^2}\frac{h\nu}{k\,T_e})$, which is negligible in the case of
magnetars and INSs. The validity of Kompaneets equation method in
the nonrelativistic case is well established (e.g. eq.(7.53) in
Rybicki \& Lightman 1979).

Let $n(\nu)$ denotes the occupation number per photon state of
frequency $\nu$. We denote the transition probability from an
initial state $(p_z,\ \nu,\ \vec{n})$ to a final state
$(p_z^{\prime},\ \nu^{\prime},\ \vec{n}^{\prime})$ as $dW$. Note
that the transition probability is a microscopic quantity, we always
have $dW^{\prime} = dW$. Since electrons move freely along the
z-direction, we describe the electron system as a 1D Maxwellian
distribution. The number of electrons with momentum in the range
$p_z-p_z+dp_z$ is $f(p_z) dp_z$, with
\begin{equation}
f(p_z)= N_e (2\pi m_e k T_e)^{-1/2} e^{-p_z^2/2m_e k T_e}.
\end{equation}
The evolution of the photon spectrum is described by the Boltzmann
equation (Rybicki \& Lightman 1979)
\begin{equation}\label{change of n}
\left( \frac{\partial n}{\partial t} \right )_{\mathrm{RCS}} = \int
dp_z \int dW [f(p_z^{\prime}) n^{\prime}(1+n) - f(p_z) n
(1+n^{\prime})].
\end{equation}
where a subscript RCS means the change of occupation number is
caused by RCS, $n$ and $n^{\prime}$ are abbreviated forms of
$n(\nu)$ and $n(\nu^{\prime})$, respectively.  For non-relativistic
electrons, the frequency change is small $\Delta \ll 1$, we can
expand eq. (\ref{change of n}) to terms of $\Delta^2$ and neglect
higher order terms. The change of photon occupation number becomes
\begin{eqnarray}\label{change of n in terms of Delta}
\left ( \frac{\partial n}{\partial t}\right )_{\mathrm{RCS}} &=&
\left [ \frac{\partial n}{\partial x} + n(1+n) \right] \int dp_z
\int dW f(p_z) \Delta \\\nonumber
 && + \left [
\frac12 \frac{\partial^2 n}{\partial x^2} + \frac{\partial
n}{\partial x}(1+n) + \frac12 n(1+n)\right] \int dp_z \int dW f(p_z)
\Delta^2.
\end{eqnarray}
We denote the two integrals by
\begin{eqnarray}
I_1 &=& \int dp_z \int dW f(p_z)
\Delta\\
I_2 &=& \int dp_z \int dW f(p_z) \Delta^2, \label{difinition of I2}
\end{eqnarray}
and then eq. (\ref{change of n in terms of Delta}) becomes
\begin{eqnarray}\label{change of n in terms of I1/I2}
\left ( \frac{\partial n}{\partial t}\right )_{\mathrm{RCS}} &=&
\left [ \frac{\partial n}{\partial x} + n(1+n) \right]
I_1\\\nonumber
 &&+ \left [
\frac12 \frac{\partial^2 n}{\partial x^2} + \frac{\partial
n}{\partial x}(1+n) + \frac12 n(1+n)\right] I_2.
\end{eqnarray}
Using the conservation of photon numbers can greatly simplify the
subsequent calculations.

The number of photons is conserved during the scattering process.
Thus we have the continuity equation of $n(x)$ in the frequency
space
\begin{equation}\label{continuity eq. vector form}
\frac{\partial n}{\partial t} = - \nabla \cdot \vec{j},
\end{equation}
where $\vec{j}$ is the photon flux in the frequency space. Assuming
$n(x)$ is isotropic (the validity of the isotropic assumption will
be discussed in the appendix), we have
\begin{equation}\label{continuity eq.}
\frac{\partial n}{\partial t} = -\frac{1}{x^2}
\frac{\partial}{\partial x} (x^2 j)
\end{equation}
A comparison between eq. (\ref{continuity eq.}) and eq. (\ref{change
of n in terms of I1/I2}) gives that the flux $j$ must have the form
(Rybicki \& Lightman 1979)
\begin{equation}\label{photon flux}
j(x) = g(x) \left[ \frac{\partial n}{\partial x} + n (1+n) \right].
\end{equation}
Note that in equilibrium conditions, $n(x) = (e^x-1)^{-1}$,
$\frac{\partial n}{\partial x} = -n(1+n)$, we have no ``photon
flux'' in the frequency space, $j=0$. This is a necessary condition.
The same condition can be used to check the validity of other forms
of diffusion equation (e.g. Liu et al. 2004). Substituting the above
equation into eq. (\ref{continuity eq.}) we have
\begin{equation}\label{continuity eq. in terms of g(x)}
\frac{\partial n}{\partial t} = -\frac{1}{x^2}
\frac{\partial}{\partial x} \left\{ x^2 g(x) \left[ \frac{\partial
n}{\partial x} + n (1+n) \right] \right\}
\end{equation}
A comparison between the coefficient of $\frac{\partial^2
n}{\partial x^2}$ in eq. (\ref{continuity eq. in terms of g(x)}) and
 eq. (\ref{change of n in terms of I1/I2}) gives g(x)
\begin{equation}
g(x) =- \frac12 I_2.
\end{equation}
Finally the Kompneets equation for RCS has the form
\begin{equation}\label{The Kompaneets eq.}
\frac{\partial n}{\partial t} = \frac{1}{x^2}
\frac{\partial}{\partial x} \left\{ x^2 \frac12 I_2 \left[
\frac{\partial n}{\partial x} + n (1+n) \right] \right\} .\nonumber
\end{equation}
We only need to calculate the integral $I_2$.

Substituting the equation of frequency change eq. (\ref{eq. of Delta
(1st term only)}) into the definition of the integral $I_2$ eq.
(\ref{difinition of I2}), first performing the integral of momentum
$p_z$, we obtain
\begin{equation}\label{integral I2}
I_2 =\int dW x^2 N_e \frac{k T_e}{m_e c^2} (\cos\theta -
\cos\theta^{\prime})^2
\end{equation}
The transition probability is directly related to the cross section
\begin{eqnarray}\label{differential cross section of RCS}
dW &=& c d\sigma_{\mathrm{RCS}}\\%
&=& c \frac{3r_e c}{32} (1+\cos^2\theta) (1+\cos^2\theta^{\prime})
\phi(\nu-\nu_B) d\Omega^{\prime}.\nonumber
\end{eqnarray}
Since we are dealing with non-relativistic electrons, the cross
section can be approximated by its value in the electron rest frame.
Performing the integral over $d\Omega^{\prime}$, we have
\begin{equation}
I_2 = 2 x^2 N_e \sigma_{\mathrm{RCS}}\, c \frac{k T_e}{m_e c^2}
g_{\theta},
\end{equation}
where $g_{\theta}$ is an angle dependent factor $g_{\theta}=\frac15
+\frac12 \cos\theta^2$. Finally the Kompaneets equation for RCS is
\begin{equation}\label{The Kompaneets eq. (final)}
\left ( \frac{\partial n}{\partial t}\right )_{\mathrm{RCS}}=
\frac{k T_e}{m_e c^2} \frac{1}{x^2} \frac{\partial}{\partial x}
\left\{x^4 N_e \sigma_{\mathrm{RCS}} \, c\, g_{\theta} \left[
\frac{\partial n}{\partial x} + n (1+n) \right] \right\}.
\end{equation}
The cross section $\sigma_{\mathrm{RCS}}$ now depends on frequency,
therefore it can not be taken out of the curly brackets. Except for
this difference and an angle dependent factor $g_{\theta}$, it is
the same as the Kompaneets equation for Compton scattering.

\section{Numerical calculations}

Before we make numerical calculations of eq. (\ref{The Kompaneets
eq. (final)}), two integrations should be performed. Performing a
space integral on both sides of eq. (\ref{The Kompaneets eq.
(final)}), employing the concept of optical depth as in eq.
(\ref{RCS optical depth}), we obtain the Kompaneets equation in the
case of pulsars
\begin{equation}\label{The Kompaneets eq. using optical depth}
\left ( \frac{\partial n}{\partial t}\right )_{\mathrm{RCS}} =
\frac{k T_e}{m_e c^2} \frac{1}{x^2} \frac{\partial}{\partial x}
\left\{x^4 \frac{\tau_{\mathrm{RCS}}}{r_1-r_2} c\, g_{\theta} \left[
\frac{\partial n}{\partial x} + n (1+n) \right] \right\}.
\end{equation}
The exprsssion $r_1-r_2$ in the denominator is the range of
integration. It is also the range of cyclotron scattering
corresponding to the frequency range $\nu_1$ to $\nu_2$. The space
integration must be employed in order to eliminate the Dirac delta
function in the RCS cross section.

There is an angular factor in the RCS optical depth
$\tau_{\mathrm{RCS}}$, we denote it as $f_{\theta}=1 +
\cos^2\theta$. To simplify the calculations, we use the average
value of $f_{\theta}$ and $g_{\theta}$
\begin{eqnarray}
\overline{f_{\theta}} &=& \frac43\\
\overline{g_{\theta}} &=& \frac25.
\end{eqnarray}
Here
$\overline{g_{\theta}}=\overline{f_{\theta}g_{\theta}}/\overline{f_{\theta}}$.
From eq. (\ref{The Kompaneets eq. (final)}) till now, we have
performed two integrations. One is integration over space range, the
other is averaging over the incoming angle. These two integrations
are introduced in order to simplify the numerical calculations.

The Kompaneets equation for RCS is a pure initial value nonlinear
partial differential equation. Solving the pure initial value
problem follows the same routine as the mixed initial value and
boundary value problem. In the case of Kompaneets equation for RCS,
we have to be careful since we are working a semi-infinite domain,
that is $0< \nu < \infty$. In the real case, we are only interested
in a finite frequency range (e.g. in the cases of magnetars, INSs).
Therefore boundary conditions are needed.

The Kompaneets equation for RCS describes the diffusion of photons
in the frequency space. It is related to the specific intensity as
\begin{equation}\label{Inu}
I_{\nu}(t)=\frac{2h\nu^3}{c^2} n(\nu,t).
\end{equation}
For a blackbody spectrum, the initial condition is\footnote{We are
considering a spherical shell of electrons, extending from $r_2$ to
$r_1$. When the radiation reaches the lower boundary $r_2$, this is
taken as $t=0$. At $r_2$, it is already several radii far away from
the neutron star surface. The gravity there is already very weak.
Therefore general relativistic effect is negligible when the
radiation propagates from $r_2$ to $r_1$.}
\begin{equation}
n(x,t=0)= \frac{1}{\exp(\frac{x}{T_{\mathrm{rad}}/T_e})-1}.
\end{equation}
Since we only consider pure scattering between electrons and
photons, the number of photons is conserved. The choice of boundary
conditions must guarantee this requirement. One guess is that there
are no photons going in or out of the specified frequency range.
Mathematically this is
\begin{equation}
\frac{\partial n(x,t)}{\partial x} + n(x,t)(1+n(x,t))=0,\quad
\mbox{when}\ x=x_1,\ x_2\quad \mbox{for all t}.
\end{equation}
Compare eq. (\ref{photon flux}) (Ross et al. 1978). During the
numerical calculations, a simplified version is used\footnote{Take
the equilibrium condition for example.
$n(\nu)=\frac{1}{e^{h\nu/kT_{\mathrm{rad}}}-1}$. At the upper
frequency boundary $h\nu_2 \gg k T_{\mathrm{rad}}$, $n\ll 1$, the
third type boundary condition is reduced to the second type. At the
lower boundary $h\nu_1 \le k T_{\mathrm{rad}}$, this is a poor
approximation. However, the number of low energy photons is
proportional to $x^2 n(x) dx \propto x^2 \ll 1$, which is also of a
small amount. Therefore we employ the simplified version of boundary
conditions. The validity of this simplified approximation is
discussed at the end of this section.}
\begin{equation}
\frac{\partial n(x,t)}{\partial x}=0, \mbox{when}\ x=x_1,\ x_2\quad
\mbox{for all t}.
\end{equation}
In obtaining the final RCS modified spectrum, we use the random walk
approximation. A photon entered the lower boundary $r_2$ escapes the
outer boundary $r_1$ after an average diffusion time scale
\begin{equation}
t_{\mathrm{dif}} = \tau_{\mathrm{RCS}}(\nu_1) (r_1-r_2)/c.
\end{equation}
In the random walk approximation, $n(x,t_{\mathrm{dif}})$ is the
final output.

\begin{figure}
\centering
  \includegraphics[width=0.6\textwidth]{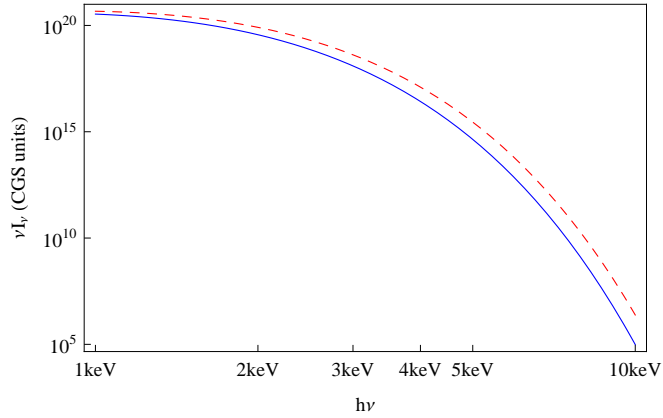}\\
  \caption{Modified blackbody spectrum due to resonant cyclotron scattering,
  up scattering case, for typical parameters of magnetars.
  The initial blackbody temperature is $0.2\, \mathrm{keV}$, the electron temperature is $10\, \mathrm{keV}$.
  The solid line is the initial blackbody spectrum.
  The dashed line is the RCS modified spectrum with optical depth $\tau_{\mathrm{RCS}}(\nu_1)=2$.
  The specified frequency range is  $(\nu_1,\nu_2)= (1\, \mathrm{keV}, 10\,\mathrm{keV})$.
  The model parameters are $(x_1, x_2)= (0.1,1)$, $(r_2,r_1)=(8.0,17)\times 10^6\,\mathrm{cm}$.
}
  \label{RCSUp}
\end{figure}

\begin{figure}
\centering
  \includegraphics[width=0.6\textwidth]{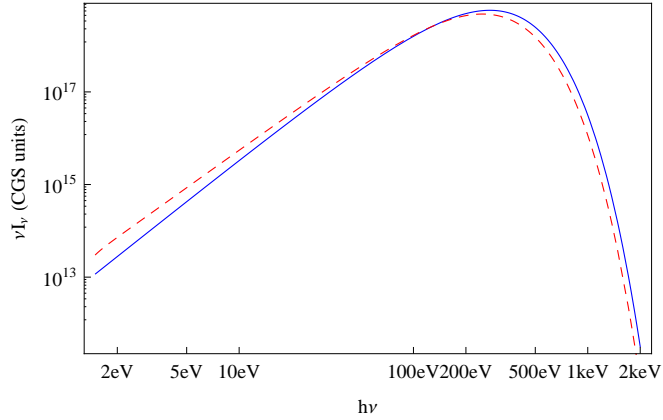}\\
  \caption{Modified blackbody spectrum due to resonant cyclotron scattering,
  down scattering case, for typical parameters of isolated neutron stars.
  The initial blackbody temperature is $70\, \mathrm{eV}$,
  the electron temperature is $26\, \mathrm{eV}$.
  The solid line is the initial blackbody spectrum.
  The dashed line is the RCS modified spectrum with optical depth $\tau_{\mathrm{RCS}}(\nu_1)=1000$.
  The specified frequency range is  $(\nu_1,\nu_2)= (1.5\, \mathrm{eV}, 2\,\mathrm{keV})$.
  The model parameters are $(x_1, x_2)= (0.058,77)$, $(r_2,r_1)=(3.9,42)\times 10^6\,\mathrm{cm}$.
  }
  \label{RCSDown}
\end{figure}

Figure \ref{RCSUp} and figure \ref{RCSDown} are the numerical
results for up scattering and down scattering case, respectively.
Figure \ref{RCSUp} shows the result for typical parameters of
magnetars. During the calculations, the magnetic field and stellar
radius are chosen as typical values of $B= 4.4\times 10^{14}\,
\mathrm{G}$ and $R= 10^6\, \mathrm{cm}$. It can reproduce a
stiffened blackbody spectrum, therefore it may be applied to
interpret magnetar soft X-ray spectrum. Figure \ref{RCSDown} shows
the calculation for typical parameters of INSs with $B= 10^{13}\,
\mathrm{G}$ and $R=10^6\, \mathrm{cm}$. It produces a spectrum with
optical/UV excess. Therefore it may account for the optical/UV
excess of INSs.

The luminosity of photons\footnote{luminosity=flux$\times$ area=$\pi
\frac{I_{\nu}}{h\nu} \left(\frac{R}{r}\right)^2 4\pi r^2$, then
integrate over the specified frequency range.}
 (number of photons per unit time passing a
fixed surface) is proportional to $\int_{x_1}^{x_2} x^2 n(x,t)\,
dx$. We can calculate this integral before and after the scattering
to check whether the number of photons is conserved during the
calculations. In the down scattering case, the specified frequency
range spans about three orders of magnitude. The number of photons
changes less than five percent before and after the scattering.
While in the up scattering case, the specified frequency range only
spans one order of magnitude. If we insert boundary conditions at
$x_1$ and $x_2$, the number of photon is not conserved. It is
because the specified frequency range is not wide enough. Therefore
we insert boundary conditions ``far away'' from the specified
frequency range, at $x_1/10$ and $15\, x_2$. The number of photons
changes less than one thousandth.

\section{Application to Isolated Neutron Stars}

ROSAT discovered seven radio quiet INSs (Kaspi et al. 2006;
Tr\"{u}mper 2005; for recent reviews see Haberl 2007; van Kerkwijk
\& Kaplan 2007). They all show featureless blackbody spectra, with
low pulsation amplitude, and high X-ray to optical flux ratio. Their
spectral energy distributions show that many of them have an
optical/UV excess with a factor of several (Burwitz et al. 2001;
Burwitz et al. 2003; Motch et al. 2003; Ho et al. 2007; van Kerkwijk
\& Kaplan 2007). Table \ref{2BB} shows double blackbody fit of RX
J1856.5-3754 (J1856 for short) and RX J0720.4-3125 (J0720 for
short). In interpreting their optical/UV excess, the emission radius
is either too small or too big for reasonable star radius. Several
theoretical models have been proposed (Motch et al. 2003; Ho et al.
2007; Tr\"{u}mper 2005). Considering the discrepancy between current
theory and observations, we try to provide an alternative one, in
which the optical/UV excess of INSs may due to magnetospheric
processes, i.e. due to down scattering of RCS when the surface
emission passing through the pulsar magnetosphere.

\begin{figure}
\centering
  \includegraphics[width=0.6\textwidth]{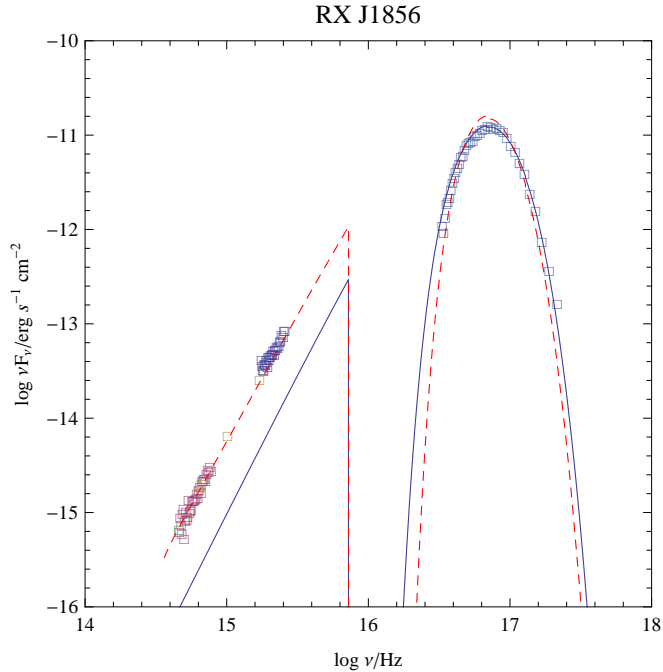}\\
  \caption{Spectral energy distributions of J1856.
  The squares are observational points (only central values are included).
  The solid line is single blackbody fit to the X-ray data, the parameters are given in table \ref{2BB}.
  (It is not the initial blackbody spectrum in our model, the parameters
  are listed in table \ref{RCSfit}.)
  The dashed line is the RCS modified blackbody spectrum, the model parameters are given in table \ref{RCSfit}.
  The specified frequency range and model parameters are the same as that in figure \ref{RCSDown}.
  All observational data are from van kerkwijk \& Kaplan (2007).
}
  \label{1856}
\end{figure}

\begin{figure}
\centering
  \includegraphics[width=0.6\textwidth]{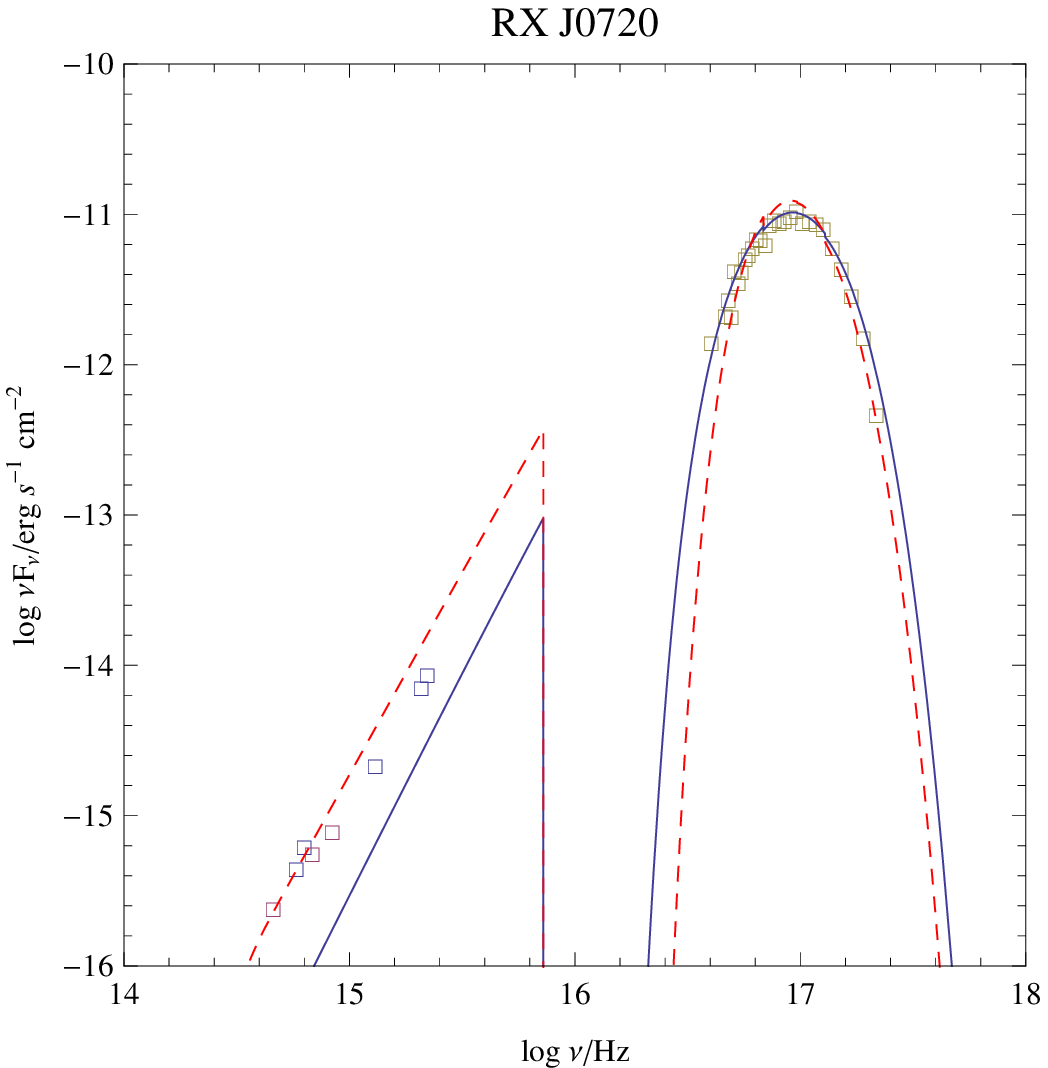}\\
  \caption{Spectral energy distributions of J0720.
  The optical/UV data of J0720 is more likely to be nonthermal.
  The specified frequency range is the same as that in figure
  \ref{RCSDown}.
  The model parameters are $(x_1, x_2)= (0.042,56)$, $(r_2,r_1)=(3.9,42)\times
  10^6\,\mathrm{cm}$.
  (The space range is the same as that in figure \ref{RCSDown},
  since it is only determined by the specified frequency range, see eq.(\ref{rrcs})).
  All observational data are from van kerkwijk \& Kaplan (2007).
  }
  \label{0720}
\end{figure}

Figure \ref{1856} and figure \ref{0720} show the RCS modified
blackbody spectrum in the case of J1856 and J0720, respectively. It
can account for the optical/UV excess in J1856 and J0720 quite well.
We assume that the initial spectrum is blackbody. When passing
through the pulsar magnetosphere, it is modified by the RCS process.
Therefore the final observed spectrum is a modified blackbody, with
optical/UV excess due to down scattering of RCS. Four parameters are
needed: the initial body temperature, the temperature of the
electron system, the electron number density (assuming homogenous),
and a normalization constant. We want to point out that our model
has the same number of free parameters as the double blackbody fit
(two temperatures and two normalization constants). The RCS model
parameters are given in table \ref{RCSfit}.

During the fitting process, the magnetic field and stellar radius
are chosen to be typical values $10^{13}\, \mathrm{G}$ and
$10\,\mathrm{km}$, respectively (e.g. Haberl 2007). In the case of
J1856, the initial blackbody temperature is chosen slightly lower
than the high temperature component of the double blackbody fit. The
temperature of the electron system is chosen as the low temperature
component of the double blackbody fit and kept fixed during the
fitting process. The electron number density are assumed to be
homogenous. The corresponding optical depth is about 1000. The
normalization constant is the solid angle of the source seen by the
observer. The photoelectric absorption cross section is from
Morrison \& McCammon (1983) (the wabs model). The hydrogen column
density is consistent with previous studies (Burwitz et al. 2003; Ho
et al. 2007).

The case of J0720 is similar. This may in part reflect the
similarities between these two INSs.  The optical/UV excess of INSs
in our model is due to magnetospheric processes. This means that, in
our model, the central star can be a normal neutron star, although
other possibilities, e.g. a quark star, can't be ruled out (Xu 2002,
2003; for a review, see Xu 2009).

In our model about the INSs, the blackbody spectrum is from the
whole stellar surface. When passing through the CFLRs of the pulsar
magnetosphere, the photons are down scattered by the RCS process. It
will bring a decrease of high energy photons. This can account for
the observed optical/UV excess. At the same time, since the X-ray
emission is from the whole stellar surface, it can also explain the
low pulsation amplitude naturally.

\begin{table}
  \centering
  \caption{Double Blackbody Fit to INS Spectral Energy Distributions.
  $T_{\mathrm{X}}$ is the high temperature component (X-ray), $T_{\mathrm{O}}$ is the low temperature component (optical/UV), seen at infinity.
  $R_{\mathrm{X}}$ and $R_{\mathrm{O}}$ are the corresponding emission radius, seen at infinity.
  Here all numbers are only estimated values, no error bars are
  given, which are taken from van Kerkwijk \& Kaplan (2007).
   }
  \label{2BB}

  \begin{tabular}{cccccc}
    \hline\hline
     & $kT_{\mathrm{X}}$ & $R_{\mathrm{X}}$ & $kT_{\mathrm{O}}$ & $R_{\mathrm{O}}$ & distance \\
    & $\mathrm{eV}$ & $\mathrm{km}$ & $\mathrm{eV}$ & $\mathrm{km}$ & pc
    \\\hline
    J1856 & 63 & 5.9 & 26 & 24.3 & 161 \\\hline
    J0720 & 85.7 & 5.7 & 35.4 & 23.5 & 330 \\
    \hline
  \end{tabular}
\end{table}

\begin{table}
  \centering
  \caption{RCS Fit to INS Spectral Energy Distributions.
  $T_{\mathrm{rad}}$ is the initial blackbody temperature, $T_{\mathrm{e}}$ is temperature of
  the electron system, seen at infinity.
  $N_{\mathrm{e}}$ is the electron number density, ``norm'' is the solid angle of the source seen by the observer.
  Equivalent hydrogen column density $n_{\mathrm{H}}$ is the result of
  wabs model.
   }
  \label{RCSfit}

  \begin{tabular}{cccccc}
    \hline\hline
     & $k T_{\mathrm{rad}}$ & $k T_{\mathrm{e}}$ & $N_{\mathrm{e}}$ & norm & $n_{\mathrm{H}}$
     \\
      & $\mathrm{eV}$ & $\mathrm{eV}$ & $10^{12}\, \mathrm{cm}^{-3}$ &  & $10^{20}\, \mathrm{cm}^{-2}$
     \\\hline
    J1856 & 61 & 26 & 1.8 & $\pi(10\,\mathrm{km}/161\mathrm{pc})^2$ & 1.6 \\\hline
    J0720 & 80 & 35.4 & 1.8 & $\pi(10\,\mathrm{km}/330\mathrm{pc})^2$ & 2.8 \\
    \hline
  \end{tabular}
\end{table}

\section{Discussions}

We consider the RCS process in this paper. Previous work of Ruderman
(2003) is mainly qualitative consideration. While our paper is a
quantitative one and detailed comparison with observational data is
also presented. Lyutilov \& Gavriil (2006) consider the RCS process
by solving the radiation transfer equation directly. We want to
point the differences between our paper and their's.
\begin{enumerate}
    \item An improved approximation is used. The photon system is modeled three
    dimensionally in our paper.
    \item A different method is employed. While Lyutikov \&
    Gavriil solved the radiation transfer equation directly, we employed
    the Kompaneets equation method. They are independent methods. The
    Kompaneets equation method is much simpler both analytically
    (compared to solving radiation transfer equation directly,
    this is why we can employ better approximations)
    and numerically (compared to doing Monte Carlo simulations).
    \item Different applications are considered. Previous researchers on
    RCS mainly focus on its application to
    magnetars. We point out that it may also play an important role in the
    radio quiet INS case. Therefore, a magnetospheric model is
    presented for the optical/UV excess of INSs.
    \item Our calculations show that there exist not only up
    scattering but also down scattering of RCS process.
\end{enumerate}

Further Monte Carlo simulations may tell us more about the down
scattering of RCS if there exists real down scattering of RCS. The
approximation of a 1D treatment (Lyutikov \& Gavriil 2006) may
result in a negative point of down scattering of RCS. Our deduction
is in the 3D case (for the photon system). The key difference is the
phase space volume (in 1D case $\propto p$, in 3D case $\propto
p^3$). While in eq. (\ref{change of n}), we note a factor ($1+n$)
appears. This is a pure second quantization effect. The photons are
aligned to condensate in the low energy sate, and this quantum
effect can only play an important role in 3D case. Note that there
is also no Bose-Einstein condensation in the low dimension case (see
Pathria 2003).

Through this paper we talk about the RCS process in pulsar
magnetospheres. We think it may be a common process. In different
cases, it has different manifestations. In the case of magnetars, we
have observed a stiffened blackbody spectrum. In the case of INSs,
we have observed an optical/UV excess. These different
manifestations can be treated universally using the Kompaneets
equation for RCS presented in this paper.

Concerning the magnetospheric properties, now people are thinking
about that the CFLRs of pulsar magnetospheres are not dead but
filled with dense plasma (Ruderman 2003; Luo \& Melrose 2007;
Lyutikov 2008). The plasma can be $10^4-10^5$ times denser than the
local Goldreich-Julian density. The origin of this over dense plasma
is the presence of twisted magnetic field lines (in the case of
magnetars) or magnetic mirroring (in the case of RRATs). When
discussing the magnetospheric properties we have to be careful. As
stated in section two, the scattering radius and the optical depth
(or cross section) is frequency-dependent. The local
Goldreich-Julian density is proportional to the magnetic field. At
the scattering sphere, from eq.(\ref{nuB}), it is proportional to
the photon frequency. Given that the electron density is a constant,
the ratio of $N_e/n_{\mathrm{GJ}}$ varies with frequency as $\propto
1/\nu$. In the case of magnetars (photon energy ranges between $1\,
\mathrm{keV}- 10\, \mathrm{keV}$), the electron density is
$10^3-10^4$ times the local Goldreich-Julian density. While in the
case of INSs, we have a broader frequency range $1\, \mathrm{eV}-
1\, \mathrm{keV}$ and a much higher RCS optical depth, about $1000$.
The corresponding electron density is $10^3-10^6$ times the local
Goldreich-Julian density. Therefore, a plasma with number density
$10^4-10^5$ times the local Goldreich-Julian density is presented in
the CFLRs of magnetars/INSs according to our model. We have computed
the mass of this dense plasma. Assuming an electron-ion plasma, the
total mass is $10^{11}\, \mathrm{g}$ and $10^{12}\, \mathrm{g}$ in
the magnetar case and INS case, respectively. This is consistent
with studies in the double pulsar binary PSR J0737-3039, Crab giant
pulses, and magnetar spectrum modeling (Lyutikov 2008; Rea et al.
2008).

The presence of dense plasmas in  CFLRs of INSs needs further
explanations. Unlike the case of magnetars, the INSs are believed to
be dead NSs (Kaspi et al. 2006; Tr\"{u}mper 2005). For slow rotators
like INSs, the magnetic mirroring mechanism comes to work (Luo \&
Melrose 2007). Therefore, we think that the dense plasma in the case
of INSs could be due to magnetic mirroring mechanism. The source of
this dense plasma may be the result of accretion from circumpulsar
material, e.g. ISM, fallback disk etc. Unlike the case of RRATs, in
the case of INSs the radiation belts are not very far away from the
neutron stars (about 40 stellar radii at the outer edge). We may
call it the ``inner radiation belt'' of a pulsar if we call the
radiation belt near the light cylinder proposed by Luo \& Melrose
(2007) the ``outer radiation belt'' of a pulsar. Nevertheless, the
particle processes are similar. The pulsar accretes material from
the environment which will be accelerated in the ``dormant outer
gap" (Luo \& Melrose). High energy curvature photons will collide
with surface X-ray photons generating pairs in INS CFLRs. The pair
plasma will be confined by the magnetic mirroring mechanism. This is
the pulsar ``inner radiation belt" (or ``electron blanket", e.g.
Ruderman 2003). Semi-quantitative estimates are given in Luo \&
Melrose (2007), Ruderman (2003). It can be as high as $10^4-10^5$
times the Goldreich-Julian density. The plasma is cold since it has
undergone a long time of relaxation (INSs are old thermally emitting
neutron stars). Meanwhile during the scattering process, the photons
will push the plasma particles away from the star. The kinetic
energy of particle decreases thus resulting in a low temperature.
This also explains why the plasma system in INS CFLRs is distributed
in a rather wide space range, see caption of figure \ref{RCSDown}.
Similar process is also possible in corona of magnetars (Beloborodov
\& Thompson 2007).

The last but not least important question is: can a neutron star
have a blackbody spectrum which can be modified when passing through
its magnetosphere? It might not be impossible. The current neutron
star atmosphere models leave us two questions: one is that a
blackbody spectrum fits the observation better than that with
spectra lines (Ho et al. 2007). The other is that we haven't found a
high energy tail in INS X-ray spectra (van Kerkwijk \& Kaplan 2007).
Therefore from the observational point of view, a blackbody spectrum
is possible. A blackbody-like spectrum could be reproduced in a
quark star model (Xu 2009).

\section{Conclusions}

We consider the RCS process in pulsar magnetospheres. The photon
diffusion equation (Kompaneets equation) for RCS is presented. It
can produce not only up scattering but also down scattering
depending on the parameter space. Its possible applications to
magnetar soft X-ray spectrum and INSs are point out.

The application to INSs is calculated in detail. We show that the
optical/UV excess of INSs may be due to down scattering of RCS. The
RCS model has the same number of parameters as the double blackbody
model. Mean while, it has a clear physical meaning. The initial
blackbody spectrum from the stellar surface is down scattered by the
RCS process when passing through its magnetosphere. This can account
for the optical/UV excess of INSs. The low pulsation amplitude of
INSs is a natural consequence in our model.

The calculations for RX J1856.5-3754 and RX J0720.4-3125 are
presented and compared with their observational data. The model
parameters for RX J1856.5-3754 and RX J0720.4-3125 are similar. This
may in part reflect the similarities between these two INSs.
Finally, we point out that the quark star hypothesis (e.g. Xu 2002)
can still not be ruled out.

The photon diffusion equation (Kompaneets equation) for RCS is
calculated semi-analytically. The calculations for the magnetar and
INS cases are all for surface thermal emission. Of course, its
application is not limited to the thermal emission case.

Using the Kompaneets equation (both the resonant and non-resonant
ones, or a unified one which will be presented in the future), a
thorough and quantitative study of scattering process in pulsar
magnetospheres could be possible. This can help us make clear the
physical process in CFLRs of pulsar magnetospheres.

\section*{Acknowledgments}

The authors would like to thank van Kerkwijk very much for providing
the observational data. H.T. would like to thank Yue You Ling, Liu
Dang Bo for helpful discussions. H.T. would like to thank Prof. Chou
Chih-Kang very much for sharing his manuscript which is also on
Kompaneets equation for resonant cyclotron scattering. H.T. and
Q.H.P. are supported by NSFC (0201131077) and the Doctoral Program
Foundation of State Education Commission of China. R.X.X. is
supported by NSFC (10935001, 10973002), the National Basic Research
Program of China (Grant 2009CB824800), and by LCWR (LHXZ200602).
L.M.S. is supported by NSFC (10778604, 10773017).

\appendix

\section{Propagation effect}

The validity of isotropic assumption employed in the main body is
discussed. From eq.(\ref{continuity eq. vector form}) to
eq.(\ref{continuity eq.}), we employ the isotropic assumption. It is
also implicitly assumed during the angular average in the numerical
calculation section. Its validity is acceptable in regions not far
away from the star. This is just the case at the inner edge $r_2$ of
the ``electron blanket". However, at the outer edge $r_1$, its
validity needs further confirmation. Our approach to this problem is
that we consider an isotropization process. From eq. (\ref{dsigma}),
the angular dependence of the outgoing photons is
$(1+\cos^2\theta^{\prime})$, the same as that of cyclotron
radiation. It is almost isotropic. Therefore, weakly dependent of
the angle of incoming photons, the photons become isotropic through
the RCS process. This provides the required isotropic photon field
to be scattered by electrons nearby. The upper isotropization
process is valid from one space location to another. Therefore, the
isotropic assumption is valid through the whole space range, from
$r_2$ to $r_1$.

Noting that the number of photons are conserved during the
scattering process. The $r^{-2}$ dependence of the solid angle of
the star is ``transformed" to the photon occupation number $n(x,t)$.
It will only modify the space integrated form of Kompaneets
equation, eq.(\ref{The Kompaneets eq. using optical depth}).
However, since the output is not sensitive to where we introduce the
$r^{-2}$ dependence, the results should be similar. Detailed
calculation is presented in below.

The energy density of the radiation field is
\begin{equation}\label{unu 1}
u_{\nu}=\frac{4\pi}{c} J_{\nu},
\end{equation}
where $J_{\nu}$ is the mean intensity
\begin{equation}
J_{\nu}=\frac{1}{4\pi}\int I_{\nu} d\Omega,
\end{equation}
$I_{\nu}$ is the specific intensity. From the definition of photon
occupation number, we have
\begin{equation}\label{unu 2}
u_{\nu}=n(\nu) \frac{8\pi \nu^2}{c^3} h\nu,
\end{equation}
where $n(\nu)$ is the photon occupation number in the Kompaneets
equation. Combining eq. (\ref{unu 1}) and (\ref{unu 2}), we obtain
the relation between mean intensity and occupation number
\begin{equation}
J_{\nu}=\frac{2h\nu^3}{c^2} n(\nu).
\end{equation}
For an isotropic radiation field (e.g. as we have assumed in the
main text), this is just eq.(\ref{Inu}). We consider the propagation
effect for a uniformly bright sphere with brightness $B_{\nu}$ and
radius $R$ (e.g., Rybicki \& Lightman 1979, section 1.3). The mean
intensity at radius $r$ is
\begin{equation}
J_{\nu} =\frac{1}{2} B_{\nu} (1-\sqrt{1-(R/r)^2}).
\end{equation}
The dilution factor is
\begin{equation}
\frac{J_{\nu}(r\gg R)}{J_{\nu}(r=R)} = \frac12 \left( \frac{R}{r}
\right)^2.
\end{equation}
This is also the dilution factor for the occupation number $n(\nu)$.

Only considering the propagation effect, it will introduce a spatial
dependence of the photon occupation number
\begin{equation}\label{r dependence of n}
\frac{\partial}{\partial r} r^2 n(\nu,r) =0.
\end{equation}
For neutron star with surface temperature $T_{\mathrm{rad}}$, the
photon occupation at radius $r$ is
\begin{equation}\label{n}
n(\nu,r\gg R) =\frac14 \frac{1}{e^{h\nu/kT_{\mathrm{rad}}}-1} \left
( \frac{R}{r} \right)^2.
\end{equation}
Besides the dilution factor, it means that only half of the photons
will propagate towards the observer.

The photon occupation number now in eq. (\ref{The Kompaneets eq.
(final)}) is a function of frequency, time, and position
$n=n(\nu,t,r)$. In order to include the dilution effect, we
introduce another variable $m$
\begin{equation}\label{m}
m(\nu,t)=r^2 n(\nu,t,r).
\end{equation}
From eq. (\ref{r dependence of n}), $m$ only depends on frequency
and time. Multiple $r^2$ on both sides of eq. (\ref{The Kompaneets
eq. (final)}) and performing a spatial integral from $r_2$ to $r_1$,
we obtain
\begin{eqnarray}\label{The Kompaneets eq. (final), after spatial integration}
\left ( \frac{\partial m}{\partial t}\right )_{\mathrm{RCS}} &=&
\frac{k T_e}{m_e c^2} \frac{1}{x^2} \frac{\partial}{\partial x}
 \left\{ x^4\, \frac{\tau_{\mathrm{RCS}}}{r_1-r_2} c\, g_{\theta}
\left[ \frac{\partial m}{\partial x} + m(1+
\frac{m}{r_{\mathrm{RCS}}^2}) \right] \right\}.
\end{eqnarray}
It is similar to eq. (\ref{The Kompaneets eq. using optical depth}).

In order to make a comparison with the observational data, the flux
of such a system must be calculate. It is related to the energy
density
\begin{eqnarray}
F_{\nu} &=& u_{\nu}c\\\nonumber
        &=& \frac{8\pi h \nu^3}{c^2} n(\nu,t,r)\\\nonumber
        &=& \frac{8\pi h \nu^3}{c^2} \frac{m(\nu,t)}{D^2}.
\end{eqnarray}
Here $D$ is distance of this source. The spectrum is proportional to
$\nu^3m(\nu,t)$. Similar results are obtained as in the main text,
with similar input parameters.

There is a deep reason why the isotropic assumption is still valid
in regions far away from the star. From eq. (\ref{rrcs}), low energy
photons will be scattered in outer regions. At the same time, they
have large cross section and optical depth, e.g. see eq. (\ref{RCS
optical depth}) and the caption of figure \ref{RCSDown}. Low energy
photons will encounter strong scattering, although they are limited
in a narrow beam. Related issue has already been pointed out in
section two.

%\newpage

\label{lastpage}

\end{document}